\documentclass[10pt]{IEEEtran}

\usepackage{float}
\usepackage{subfigure}
\usepackage{stfloats}
\usepackage{cite}
\usepackage{amsmath,amssymb,amsfonts}
\usepackage{graphicx}
\usepackage{textcomp}
\usepackage{mathrsfs}
\usepackage{booktabs}
\usepackage{cases} 
\usepackage{algorithm}
\usepackage{algorithmicx}
\usepackage{algpseudocode}
\usepackage{color}
\usepackage{amsmath}

\newcommand{\be}{\begin{equation}}
	\newcommand{\ee}{\end{equation}}
\def\bee#1\eee{\begin{align}#1\end{align}}
\newcommand{\bse}{\begin{subequations}}
	\newcommand{\ese}{\end{subequations}}

\ifCLASSINFOpdf
\else
\fi

\begin{document}
	
	\title{Technical Report for Trend Prediction Based Intelligent UAV Trajectory Planning for Large-scale Dynamic Scenarios}

	\author{Jinjing Wang, Xindi Wang}
	\maketitle
	
	\begin{abstract}
		The unmanned aerial vehicle (UAV)-enabled communication technology is regarded as an efficient and effective solution for some special application scenarios where existing terrestrial infrastructures are overloaded to provide reliable services. To maximize the utility of the UAV-enabled system while meeting the QoS and energy constraints, the UAV needs to plan its trajectory considering the dynamic characteristics of scenarios, which is formulated as the Markov Decision Process (MDP). To solve the above problem, a deep reinforcement learning (DRL)-based scheme is proposed here, which predicts the trend of the dynamic scenarios to provide a long-term view for the UAV trajectory planning. Simulation results validate that our proposed scheme converges more quickly and achieves the better performance in dynamic scenarios.
	\end{abstract}

	
	\IEEEpeerreviewmaketitle

	\section{Introduction}
	Unmanned Aerial Vehicles (UAVs) have been used in providing emergency communication services for some special scenarios that cannot be served satisfactorily with existing terrestrial infrastructures, such as some mass gathering events, like new year events, large conferences, etc. In these situations, the trajectory planning of the UAV involved can seriously affect system performances, e.g., data throughput\cite{Sun2019}, transmission delay\cite{MengTVT2021}, and user fairness\cite{Liu2018}. 
	
	For the practical applications, the classical optimization-based UAV trajectory planning strategies \cite{WangCL2021} may no longer be feasible, as most of them are designed to be operated in an iterative fashion with high computational complexity. In the light of the aforementioned, Reinforcement Learning (RL) \cite{LiuTMC2021} based strategies  have been regarded as promising solutions for the UAV-enabled system owing to the great self-learning capability, which exhibits significant efficiency gains in static scenarios by optimizing the interaction process \cite{Seid2021} and experience selection \cite{Li2022}. 
	Furthermore, faced with realistic application scenarios with highly dynamic characteristics, more and more researches have studied on how to extract scene features by adding Deep Neural Networks to further enhance the capability of RL, so as to construct the Deep RL framework (DRL). To achieve above goals, \cite{Oubbati2021,Luong2021,Qin2021,Liu2020} construct the Artificial Neural Network (ANN) to mine the correlations between scene information (e.g. location \cite{Oubbati2021,Luong2021}, energy state \cite{Luong2021} and throughput \cite{Qin2021,Liu2020}). In order to avoid the uncertainty caused by manual feature extraction in ANN-based schemes, \cite{LiuTMC2021,Huang2020,ZhangTAS2019} construct the Convolutional Neural Network (CNN) and model the scene information as a tensor composed of multiple channels (e.g. communication range \cite{LiuTMC2021}, location \cite{Huang2020} and object tracking information \cite{ZhangTAS2019}), which makes scene features more hierarchical.
	However, considering the dynamic characteristics of real scenarios, there are still two major research gaps in existing researches: 1) The number of GUs is related to the structure of the DRL framework\footnote{Assuming that each GU has $n$ features, then the input of the ANN-based model is a vector of dimension $nN\times 1$ where $N$ denotes the number of GUs in the scenario. Apparently, a change of $N$ can lead to changes in the network structure.} designed in existing studies \cite{Oubbati2021,Luong2021,Qin2021,Liu2020}, the variation in the number of GUs in dynamic scenarios leads to the need for re-tuning and re-training for the proposed DRL models. 2) The scene information of GUs is not fully utilized in existing studies \cite{LiuTMC2021,Huang2020,ZhangTAS2019} where the whole dynamic  process is usually processed as multiple independent frames, which lacks further exploration of their associations.
	
	Therefore, the critical issue is how to bridge the above gaps by designing a DRL framework that is flexible enough and can fully exploit the dynamic characteristics of the scene, which motivates this work with the objective to maximizes the long-term performance of the UAV-enabled large-scale dynamic scenarios by optimizing the UAV's movement \textbf{action}, subjected to the constraints of the communication QoS and energy. Therefore, we design a moving Trend Prediction (TP) based DRL framework for the UAV (served as an \textbf{agent}) to perceive the \textbf{state} of the current environment and predict the trend of the future \textbf{state}. Through continuous interaction with the environment, the \textbf{agent} optimizes its actions according to the received feedback (known as \textbf{reward} in DRL). Simulations have been used to verify and validate the performance of the proposed scheme.
	
	\section{System Model and Problem Formulation}
	\subsection{System Model}
	Consider an area of interest (AoI) where the explosive growth of access requirements have already far outstripped the capacity of existing terrestrial base stations. One UAV with velocity $v_{uav}$ is dispatched to provide the extra communication capacity for the GUs inside the AoI (denoted as a GU set $\Omega_{all}^t$) with fixed flight altitude $H$. In particular, the entire mission period of the UAV is discretized into multiple individual time slots and each with equal duration $\tau$. Similarly, referring to \cite{Liu2020}, the AoI has been divided into $K\times K$ equal grids, of which the centers are used as way-points of the UAV in each time slot. Thus, the location of the UAV (projected on the ground) and the $i$-th mobile GU in time slot $t$ can be formulated as $\mathbf{L_u^t}=[x_u^t,y_u^t]$ and $\mathbf{L_i^t}=[x_i^t,y_i^t]$, respectively. Following the model proposed in \cite{Liu2020}, the velocity and moving direction of the GU $i$ will be updated as
	\begin{align}
		v_i^t&=k_1 v_i^{t-1}+(1-k_1)\overline{v},\\
		\theta_i^t&=\theta_i^{t-1}+\widetilde{\theta}k_2.
	\end{align}
	where $\overline{v}$ and $\widetilde{\theta}$ denote average velocity and steering angle, respectively, and $k_2$ follows $\epsilon$-greedy model\footnote{In each time slot, the GU will keep the same moving direction with probability $\epsilon$, otherwise the GU chooses one of the remaining directions randomly.}. 
	
	Since the terrestrial infrastructures cannot provide communication services, the GUs' data will be temporarily stored in their on-board buffer and wait for the UAV to start the data upload process.
	Here, the GU-to-UAV channel is modeled as Rician models \cite{Sun2019} that can capture the shadowing and small-scale fading effects due to multi-path propagation, in which the channel coefficient from the GU $i$ to the UAV in time slot $t$ can be expressed as
	\begin{align}\label{eq:h}
		h_i^t=\sqrt{\beta_i^t}\hat{h_i^t},
	\end{align}
	where $\beta_i^t=\frac{\alpha}{(H^2+||\mathbf{L_u^t}-\mathbf{L_i^t}||^2)^{k_{ps}/2}}$ and $\hat{h_i^t}=\sqrt{\frac{k_s}{k_s+1}}\tilde{h_i^t}+\sqrt{\frac{k_s}{k_s+1}}\tilde{\tilde{h_i^t}}$, $|\tilde{h_i^t}|=1$, $|\tilde{\tilde{h_i^t}}|\sim CN(0,1)$, and $K_{ps}$ and $K_s$ denote the GU-to-UAV path loss component and Rician factor, respectively.                                                                                                                                                                                                                                                           
	$\alpha$ denotes the channel power gain at the reference distance $||\mathbf {L_u^t}-\mathbf{L_i^t}||=1$m. 
	
	By adopting OFDMA technology, the UAV can pre-divide communication resources into multiple equal and orthogonal resource blocks in advance, of which the bandwidth allocated to each communication GU is $W$. In this way, the UAV using OFDMA technology can support concurrent data transmissions with surrounding GUs. Thus, the communication rate with respect to each GU is given by $r_i^t=W\log_2(1+\frac{|h_i^t|^2p}{\sigma^{2}})$, where $p$, $W$, and $\sigma^2$ denote the transmission power, bandwidth and noise power, respectively. In this way, the data uploaded throughput of the $i$-th GU is obtained by $r_i^t\tau_c$, in which $\tau_c$ denotes the duration of the hovering time used for communications within one time slot, which is assumed as $\tau\gg \tau_c$. Here we denote the data queue length of the GU $i$ in time slot $t$ as $B_i^t$, which depends on both the newly generated data (i.e. $I_i^t$) and the data uploaded (i.e. $r_i^t\tau_c$) to the UAV, which is given by $B_i^t={B_i^{t-1}+I_i^t-r_i^t\tau_c}$.
	
	In addition, the energy consumption of the UAV in time slot $t$ is associated with the flying distance of the UAV in the past time slot \cite{Liu2020}, which is formulated as 
	\begin{align}\label{eq:E}
		E_f^t=p_f\frac{||\mathbf {L_u^t}-\mathbf{L_u^{t-1}}||}{v_{uav}},
	\end{align}
	where $p_f$ denotes the UAV flying power.

	\subsection{Problem Formulation}
	Regarding the dynamic scenarios considered in this work, the location, number, and data queue of GUs that change in real time greatly increases the feature dimension of the UAV-Ground communication scenario, which poses a severe challenge to build the efficient and reliable data links between GUs and the UAV. Therefore, our objective is to find a policy that helps the UAV make the optimal decision on trajectory planning, so that the utility of the UAV-enabled communication system over all the time slots is maximized, subjected to the coverage range, energy consumption, and the communication QoS. Thus, the UAV trajectory planning issue\footnote{We summarize the notations used in this paper in section \uppercase\expandafter{\romannumeral2} of our technical report \cite{Wang2021tech}.} is formulated as
	\begin{align}
		\textup{P0}:&\max\ \sum\nolimits_{t\in \mathcal{T}}\mathcal{U}^t\nonumber\\
		\ \textup{s.t.}\ &{\text{C1}}:\  h_i^t\geq \underline{h},\  \forall \ t,\nonumber\\
		&{\text{C2}}:\  \sum_{t}E_f^t\leq \overline{E},\nonumber
	\end{align}
	where the system utility function $\mathcal{U}^t$ depends on both the throughput and fairness during the data upload processes. Furthermore, C1 ensures that the communication link should meet the required QoS, namely the channel coefficient of each communication GU $i$ (see \eqref{eq:h}) should not be smaller than the threshold $\underline{h}$. C2 limits the total energy consumption during the UAV flight process $\sum_{t}E_f^t$ (see \eqref{eq:E}) to within the portable energy $\overline{E}$ under the premise that the communication energy consumption is negligible.
	
	Furthermore, the utility function $\mathcal{U}^t$ is defined here, $\mathcal{U}^t=f_G^t\sum_{i\in \Omega_{all}^t}r_i^t\tau_c$, in which $f_G^t$ called as the Jain's fairness index \cite{Liu2018} is defined as
	\begin{align}\label{eq:fair}
		f_G^t=\frac{(\sum_{i\in \Omega_{all}^t} c_i)^2}{|\Omega_{all}^t|\sum_{i\in \Omega_{all}^t} c_i^2},
	\end{align}
	in which $c_i$ denotes the number of UAV-enabled communication services that GU $i$ has participated in. As a widely-used metric for fairness, the value of $f_G^t$ approaches $1$ when the total number of time slots that each GU is served are very close, which can be regarded as a measure of the fairness of UAV communication services in existing scenarios.

	\subsection{MDP Model}
	In general, conventional methods normally would narrow down \textup{P0} for an individual time slot, namely $\max  U^t,\forall t\in \mathcal{T}$, and solve it by classical convex optimization methods or heuristic algorithms, which may obtain the greedy-like performance due to lacking of a long-term target. To tackle above issue, we apply Markov Decision Process (MDP), defined as a tuple ($\mathcal{S}$, $\mathcal{A}$, $\mathcal{R}$, $\mathcal{P}$), to model the UAV trajectory planning problem in large-scale dynamic scenarios, which are detailed as follows.
	
	1) The $state$ $\mathcal{S}$ contains the locations of the UAV and GUs as well as their real-time status, which is formulated as 
	\begin{align}
		\mathcal{S}\triangleq\{s_t=\{\mathbf{L_u^t},\mathbf{L_i^t},B_i^t,h_i^t|i\in \Omega_{all}^t, \forall t\}\}.
	\end{align}
	
	2) The $action$ $\mathcal{A}$ contains available actions of the UAV in each time slot, which is formulated as
	\begin{align}
		\mathcal{A}\triangleq&\{a_t=\{Up, Down, Left, Right,Right\ upper,\\\nonumber & Right\ lower, Left\ upper, Left\ lower|\forall t\} \}.
	\end{align}
	
	3) The $reward$ $\mathcal{V}_{\pi}(s_t)$ represents the discounted accumulated reward from the state $s_t$ to the end of the task with the policy $\pi$, which is formulated as
	\begin{align}\label{eq:V}
		\mathcal{V}_{\pi}(s_t)=\mathbb{E}_{\pi}[\sum_{j=0}^{\mathcal{J}}\gamma^j r(s_{t+j},a_{t+j})],
	\end{align}
	where $r(s_t,a_t)$ denotes the immediate reward through executing action $a_t$ at the state $s_t$, and $\mathcal{J}$ denotes the end of the task. In particularly, the action $a_t$ adopted here is selected following the policy $\pi$, i.e., $\pi(s_t)=a_t$.  According to problem P0, $r(s_t,a_t)$ is defined as 
	$$r(s_t,a_t)=\left\{
	\begin{aligned}
		\mathcal{U}^t&,&\ {\text{if C1 and C2 are satisfied,}}\\
		0&,&\ {\text{otherwise.}}
	\end{aligned}
	\right.
	$$
	
	4) The $transition$ $probability$ $\mathcal{P}\triangleq\{p(s_{t+1}|s_t)\}$ represents the probability that the UAV reaches the next state $s_{t+1}$ while in the state $s_t$, which is formulated as
	$$
	p(s_{t+1}|s_t)=\left\{
	\begin{aligned}
		&\eta\ \ \ \ , & \text{if}\  s_{t+1}\triangleq\arg\max_{a_t\in\mathcal{A}} \mathcal{V}_{\pi}(s_{t+1}|s_t,a_t), \\
		&1-\eta , & \text{otherwise},
	\end{aligned}
	\right.
	$$
	where $\eta$ corresponds to the greedy coefficient during the action selection process.
	
	Based on the formulated MDP model ($\mathcal{S}$, $\mathcal{A}$, $\mathcal{R}$, $\mathcal{P}$), the UAV first observes the state $s_t$ in each time slot $t$. Then, it takes an action $a_t$ following the policy $\pi$, i.e., $a_t=\pi(s_t)$, and thus obtains the corresponding immediate reward $r(s_t,a_t)$. Then, the UAV moves to the next way-point and the $state$ updates to $s_{t+1}$. Therefore, the problem P0 can be transformed into maximizing the discounted accumulated reward by optimizing the policy $\pi$.

	A table that summarizes all notations in this paper is given in Table.~1.
	\section{$N$-Step Mobile Trend Based DRL Scheme}
	\subsection{DRL Framework}
	It is well-known that the DRL algorithm is efficient and effective in solving MDP for the uncertain (i.e. $\mathcal{V}_{\pi}(s_t)$) and complex (i.e. $\mathcal{S}$) system, in which the critical issue here is how to obtain the optimal policy $\pi^{*}$ for the action selection process. 
	
	To achieve that, we first rephrase $\mathcal{V}^{\pi}(s_t)$ into the form of state-action pairs as $\mathcal{V}^{\pi}(s_t)=Q^{\pi}(s_t,a_t)$ where $a_t=\pi(s_t)$ means the action $a_t$ is selected at the state $s_t$ according to the policy $\pi$. Here we call $Q^{\pi}(s_t,a_t)$ as the Q-function. Referring to \cite{Liu2020}, it is easy to draw the conclusion that the optimal policy can be derived by $\pi^{*}(s_t)=\mathop{\arg\max}\limits_{a\in \mathcal{A}} Q(s_t,a)$. In other words, once the UAV selects the action that can maximize the corresponding Q-function at each $state$ $s_t\in \mathcal{S}$, the optimal policy can be realized. Then, the remaining issue is how to obtain $Q(s_t,a_t)$ with the given environment $s_t$ and action $a_t$.


	Based on above conclusions and \eqref{eq:V}, we have
	\begin{align}\label{eq:reg}
		Q^{\pi^{*}}(s_t,a_t)=r(s_t,a_t)+\gamma Q^{\pi^{*}}(s_{t+1},\pi^{*}(s_{t+1})),
	\end{align}
	where $Q^{\pi^{*}}(s_{t+1},\pi^{*}(s_{t+1}))=\mathop{\max}\limits_{a\in\mathcal{A}} Q^{\pi^{*}}(s_{t+1},a)$, thus the issue about searching the Q-function $Q^{\pi^{*}}(s_t,a_t)$ can be formulated as a regression problem, that is, through iteratively optimizing the parameters of the Q-function so that the left-hand side of the \eqref{eq:reg} is infinitely close to the right-hand side. Here, considering the high dimension of $\mathcal{S}$ and $\mathcal{A}$ in the scenario we studied, it is common to apply the neural networks (NN) to fit the Q-function mentioned above. To be specific, the NN can address the sophisticate mapping between the $state$ $s_t$, $action$ $a_t$ and their corresponding Q-function value $Q(s_t,a_t)$ based on a large training data set. Accordingly, two individual deep neural networks are established here. One with parameters $\theta_P$ is utilized to construct the \textbf{Evaluation} network $Q_{\theta_P}(s_t,a_t)$ that models the function $ Q^{\pi^{*}}(s_t,a_t)$, and another one with parameters $\theta_T$ is used to construct the \textbf{Target} network $Q_{\theta_T}(s_t,a_t)$ for obtaining the target value (i.e., $r(s_t,a_t)+\gamma Q^{\pi^{*}}(s_{t+1},\pi^{*}(s_{t+1}))$) in the training process. Finally, the parameters $\theta_P$ of $Q_{\theta_P}(s_t,a_t)$ are updated by minimizing the loss function $L$, which is formulated as
	\begin{align}\label{eq:offupdate}
		L=\mathbb{E}_{s_t\in \mathcal{S}}[(\underbrace{Q_{\theta_p}(s_t,a_t)}_{{\text{Evaluation}}} -\underbrace{(r(s_t,a_t)+\gamma\max Q_{\theta_t}(s_{t+1},a))}_{{\text{Target}}})^2].
	\end{align}
	
	Finally, after the NN has been well trained, the UAV can make a decision at the state $s_t$ according to the obtained Q-function that is modeled as the NN, and then the UAV takes the action $a_t=\arg\max_{a\in\mathcal{A}} Q_{a}(s_t,a)$
	\subsection{Design the Input Layer of Neural Network}
	In order to extract spatial features of the scenario, here we adopt Convolutional Neural Networks (CNNs) to achieve the \textbf{Evaluation} network $Q_{\theta_P}(s_t,a_t)$, which sets a three-channel tensor with size  $\mathbb{R}^{K\times K\times 3}$ as the input. Meanwhile, each channel of the tensor is modeled as a matrix with the size $\mathbb{R}^{K\times K}$, corresponding to the scenario model that is divided into $K\times K$ equal grids. (see the system model in Section.~\uppercase\expandafter{\romannumeral2}-A). Here, two convolution layers are constructed to extract the features from the input, and two full connected layers are constructed to establish associations between them. The  design of the three-channel tensor are given as follows.
	\subsubsection{Channel 1} The Channel $1$ describes the effective communication range of the UAV with the location $\mathbf{L_u(t)}$, which is formulated as a matrix $\mathbf{T_1}\in\mathbb{R}^{K\times K}$. To meet the QoS during the Ground-to-UAV communication process, we substitute (3) into the constraint C1, namely
	\begin{align}\label{eq:Ccom}
		||\mathbf{L_u(t)}-\mathbf{L_i(t)}||^2\leq (\frac{\alpha \hat{h_i(t)}^2}{\underline{h}^2})^{2/K_{ps}}-H^2.\nonumber
	\end{align}
	which indicates that the location of GUs that can establish reliable communication with the UAV must meet above condition. In this way, we can calculate whether the distance from GU $i\in\Omega_{all}$ to the location of the UAV $\mathbf{L_u(t)}$ satisfies the above condition. If so, we assign the corresponding element in matrix $\mathbf{T_1}$ as the real-time channel coefficient obtained, otherwise set it as $0$.
	
	\subsubsection{Channel 2} The Channel $2$ describes the interaction information between GUs and the UAV, which is formulated as a matrix $\mathbf{T_2}\in\mathbb{R}^{K\times K}$. In $\mathbf{T_2}$, we record the communication times between each GU $i$ and the UAV (i.e., $c_i$), and set the element of $\mathbf{T_2}$ corresponding to the location of GU $i$ as $c_i,i\in\Omega_{all}$, otherwise is $0$.
	
	\subsubsection{Channel 3} The Channel $3$ describes the predicted movement trend of GUs in the future $N$ time slots, which is formulated as a matrix $\mathbf{T_3}\in\mathbb{R}^{K\times K}$. The design principle of $\mathbf{T_3}$ is predicting the future movement situation of GUs by extracting their mobile features from historical records, which enables the UAV a future-oriented view during the trajectory planning to achieve the maximum accumulated reward from the current state until the end of the task. The specific steps for constructing the channel $3$ are as follows.
	
	$\bullet$ Step 1: The matrix $\textbf{G}^t\in \mathbb{R}^{K\times K}$ is produced to record the buffer state of GUs in the scenario, in which the element is set as $\textbf{G}^t[m,n]=B_i^t$ if GU $i\in\Omega_{all}$ is located in the corresponding grid (i.e. the $m$-th row and the $n$-th column) of the scenario, otherwise $\textbf{G}^t[m,n]=0$. 
	
	$\bullet$ Step 2: The difference matrix $\Delta \textbf{G}^t\in \mathbb{R}^{K\times K}$ is produced as $\Delta \textbf{G}^t=\textbf{G}^t-\textbf{G}^{t-1}$ here, showing the changes of $\textbf{G}^t$ between adjacent time slots.
	
	$\bullet$ Step 3: Four direction kernels $\mathcal{U}\in \mathbb{R}^{2\times 1}$, $\mathcal{D}\in \mathbb{R}^{2\times 1}$,  $\mathcal{L}\in \mathbb{R}^{1\times 2}$, and $\mathcal{R}\in \mathbb{R}^{1\times 2}$ are utilized here to detect the moving direction of GUs (corresponding to up, down, left and right respectively), which are given by
	\begin{align}
		\mathcal{U}=[1,-1]^T,\mathcal{D}=[-1,1]^T,\mathcal{L}=[1,-1],\mathcal{R}=[-1,1].
	\end{align}
	
	$\bullet$ Step 4: The SAME convolution operation is performed on the difference matrix $\Delta\textbf{G}^t$ using the above four kernels, respectively. The corresponding outputs are four matrices $\textbf{G}_{\mathcal{U}}^t$, $\textbf{G}_{\mathcal{D}}^t$, $\textbf{G}_{\mathcal{L}}^t$ and $\textbf{G}_{\mathcal{R}}^t$ with the same size $\mathbb{R}^{K\times K}$, which denotes the detection result of the movement direction of GUs in the past two time slots, respectively. Specifically, elements of above four matrices are obtained by 
	$$
	\setlength\abovedisplayskip{-0.5pt}
	\left\{
	\begin{aligned}
		\textbf{G}^t_{\mathcal{U}}[m,n]=\sum\nolimits_{i=0}^{1} \Delta \textbf{G}^t[m+i,n]\times\mathcal{U}[i], \\
		\textbf{G}^t_{\mathcal{D}}[m,n]=\sum\nolimits_{i=0}^{1} \Delta \textbf{G}^t[m+i,n]\times\mathcal{D}[i], \\
		\textbf{G}^t_{\mathcal{L}}[m,n]=\sum\nolimits_{i=0}^{1} \Delta \textbf{G}^t[m,n+i]\times\mathcal{L}[i], \\
		\textbf{G}^t_{\mathcal{R}}[m,n]=\sum\nolimits_{i=0}^{1} \Delta \textbf{G}^t[m,n+i]\times\mathcal{R}[i].
	\end{aligned}
	\ \forall m,n.\right.
	$$
	
	$\bullet$ Step 5: Based on step $4$, the trend prediction matrices of up, down, left and right directions are constructed respectively, which are denoted as $\textbf{T}^N_{\mathcal{U}}$, $\textbf{T}^N_{\mathcal{D}}$, $\textbf{T}^N_{\mathcal{L}}$ and $\textbf{T}^N_{\mathcal{R}}$.
	Taking $\textbf{T}^N_{\mathcal{R}}$ as an example, the specific steps are given in Algorithm.~1.  
	
	\begin{algorithm}
		\caption{Generate the matrix $\textbf{T}^N_{\mathcal{R}}$ with $N$ steps}
		\label{alg:sp1}
		{\small{
				\begin{algorithmic}[1] 
					\State Initialize $i=1$, generate $\textbf{T}^N_{\mathcal{R}}[m,n]$ as
					$$\textbf{T}^N_{\mathcal{R}}[m,n]=\left\{
					\begin{aligned}
						\textbf{G}_{\mathcal{R}}^t[m,n]&,&\ {\text{if}}\ \textbf{G}_{\mathcal{R}}^t[m,n]>0, \\
						0&,&{\text{otherwise}}.
					\end{aligned}
					\right.
					$$	
					\For{each $\textbf{T}^N_{\mathcal{R}}[m,n]\neq 0$,\ set $x\leftarrow m$, $y\leftarrow n$}
					\For{$i\leq N$}
					\State With probability $\epsilon$, set $\textbf{T}^N_{\mathcal{R}}[x+1,y]=\textbf{T}^N_{\mathcal{R}}[x+1,y]+\gamma\textbf{T}^N_{\mathcal{R}}[x,y]$, and $x\leftarrow x+1$. (\text{Right})
					\State Otherwise, select another direction randomly, and update the corresponding element as 
					\begin{align}
						\textbf{T}^N_{\mathcal{R}}[x-1,y]=\textbf{T}^N_{\mathcal{R}}[x-1,y]+\gamma\textbf{T}^N_{\mathcal{R}}[x,y], x\leftarrow x-1, (\text{Left})\nonumber
					\end{align}
					or \begin{align}
						\textbf{T}^N_{\mathcal{R}}[x,y+1]=\textbf{T}^N_{\mathcal{R}}[x,y+1]+\gamma\textbf{T}^N_{\mathcal{R}}[x,y], y\leftarrow y+1,(\text{Up})\nonumber
					\end{align}
					or \begin{align}
						\textbf{T}^N_{\mathcal{R}}[x,y-1]=\textbf{T}^N_{\mathcal{R}}[x,y-1]+\gamma\textbf{T}^N_{\mathcal{R}}[x,y], y\leftarrow y-1.(\text{Down})\nonumber
					\end{align}
					\State i=i+1.
					\EndFor
					\EndFor
					\State Normalize $\textbf{T}^N_{\mathcal{R}}$.
		\end{algorithmic}}}
	\end{algorithm}
	
	For $\textbf{T}^N_{\mathcal{L}}$, $\textbf{T}^N_{\mathcal{U}}$, $\textbf{T}^N_{\mathcal{D}}$, their elements are updated with the similar principle as in step $5$ of the Algorithm.~1, namely
	$$
	\setlength\abovedisplayskip{-0.5pt}
	\left\{
	\begin{aligned}
		\textbf{T}^N_{\mathcal{L}}[x-1,y]=\textbf{T}^N_{\mathcal{L}}[x-1,y]+\gamma\textbf{T}^N_{\mathcal{L}}[x,y], x\leftarrow x-1,\\
		\textbf{T}^N_{\mathcal{U}}[x,y+1]=\textbf{T}^N_{\mathcal{U}}[x,y+1]+\gamma\textbf{T}^N_{\mathcal{U}}[x,y], y\leftarrow y+1,\\
		\textbf{T}^N_{\mathcal{D}}[x,y-1]=\textbf{T}^N_{\mathcal{D}}[x,y-1]+\gamma\textbf{T}^N_{\mathcal{D}}[x,y], y\leftarrow y-1.
	\end{aligned}
	\right.
	$$

	$\bullet$ Step 6: output the $N$-step moving trend prediction matrix $\mathbf{T_3}=\textbf{T}^N_{\mathcal{U}}+\textbf{T}^N_{\mathcal{D}}+\textbf{T}^N_{\mathcal{L}}+ \textbf{T}^N_{\mathcal{R}}$, and set it as the Channel $3$ of the CNN model.
	
	\subsection{Joint Offline and Online Learning Framework}\label{sec:C}
	In order to enable the UAV to update the DRL model at the same time during executing the communication task, a training framework that combines online and offline manner is proposed, in which the UAV performs the task following the policy obtained at the offline stage, and then updates the its policy online in current episode.
	
	The offline learning process is designed to learn the most valuable experiences from the past. Therefore, a large "Replay Memory" (denoted as $R_l$) is constructed to store the experiences including $state$ $s_t$, $action$ $a_t$, and reward $\mathcal{U}^t$ during the past episodes, of which seventy percent are with the largest reward and the remaining thirty percent are chosen randomly from the rest. In contrast, the online learning process aims to learn from the current experiences, resulting in the current best behavioral decisions. Therefore, it will generate a relative small "Replay Memory" (denoted as $R_s$) to store the experiences sampled during the current communication task, of which eighty percent are with the largest reward from the current episode and the remaining twenty percent are picked randomly. 
	
	The combination of the above two training processes not only allows the UAV to take full advantage of past experiences, but also adjust its policy according to the current actual situation, and sum up all past experiences after the end.
	
	\subsection{Computational Complexity}
	The computational complexity of our proposed moving Trend Prediction (TP) based DRL framework is mainly involved in two parts, namely the building process of the channel $3$ and the CNN calculation process. The former can be obtained as follows: step $1$, $2$ and $3$ only involve simple calculations, step $4$ requires $O(4\times|\Omega_{all}^t|)$ and step $5$ requires $O(8\times N\times|\Omega_{all}^t|)$. Then, the computational complexity of the CNN calculation process is given by $O(N_c\times M_c^2\times K_c^2\times C_{in}\times C_{out})$ \cite{Chollet2017} in which $N_c$=$2$ $M_c$=$32$, $K_c$=3, $C_{in}$=2 and $C_{out}$=32 are the parameters of the model in our scheme. 
	
	\section{NUMERICAL RESULTS}
	The simulation parameters of the scenario are set as $|\Omega_{all}^t|=50$, $K=30$, $k_1=0.9$, $\overline{v}=1$m/s, $v_{uav}=30m/s$, $p_f=110$W, $\widetilde{\theta}=\frac{\pi}{2}$, $H=40$m, $K_{ps}=2$, $K_s=1$ and $k_2$ follows $\epsilon$-greedy model with $\epsilon=0.9$.  The simulation parameters of the communication process are set as $W=2$Mhz, $\tau=1$s, $\tau_c=0.1$s, $\underline{h}=2.5\times 10^{-9}$, $\overline{E}=10^4$kJ, $p=0.1$W $\alpha=10^{-5}$, $\sigma=10^{-9}$ and $I_i^t=5\times 10^{-3}$bits/s. The Trend Prediction based model proposed in this work is called as TP here, in which the greedy coefficient $\eta=0.9$. The results are shown in Fig.~1 and Fig.~2.
	
	\begin{figure}[htbp]
		\centering
		\subfigure[]{
			\includegraphics[width=0.4\textwidth]{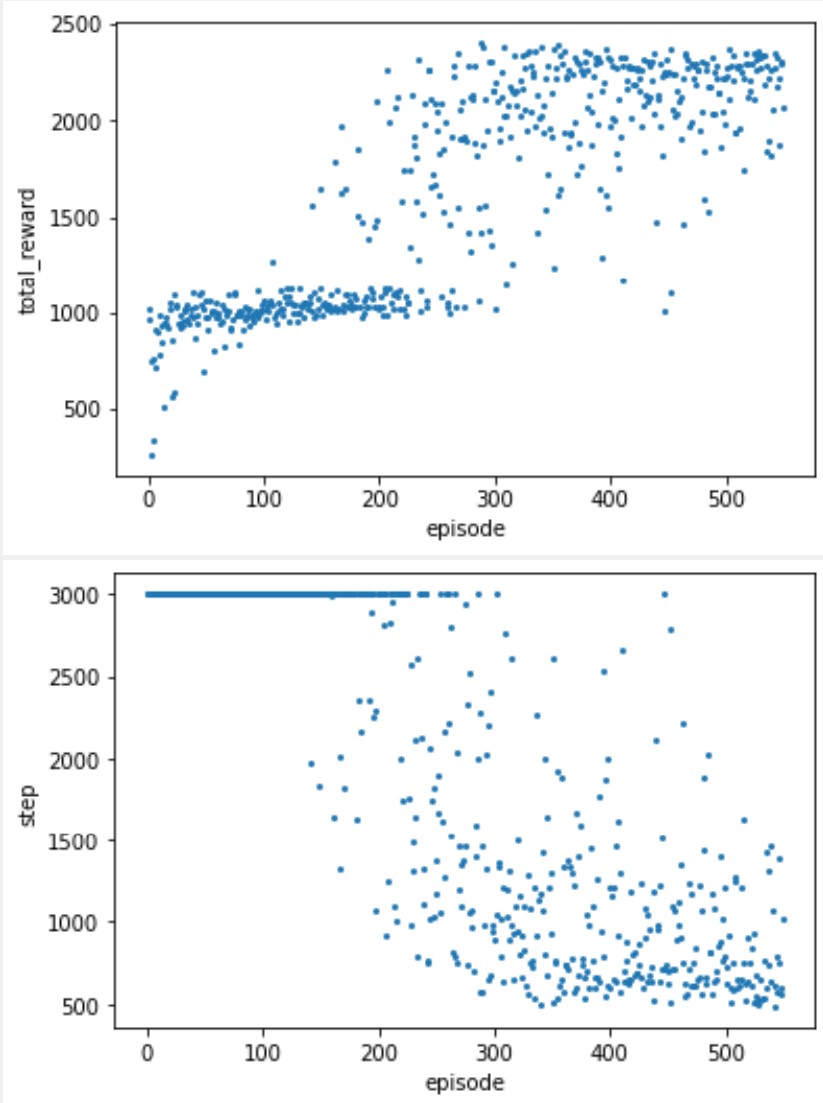}
		}
		\subfigure[]{
			\includegraphics[width=0.4\textwidth]{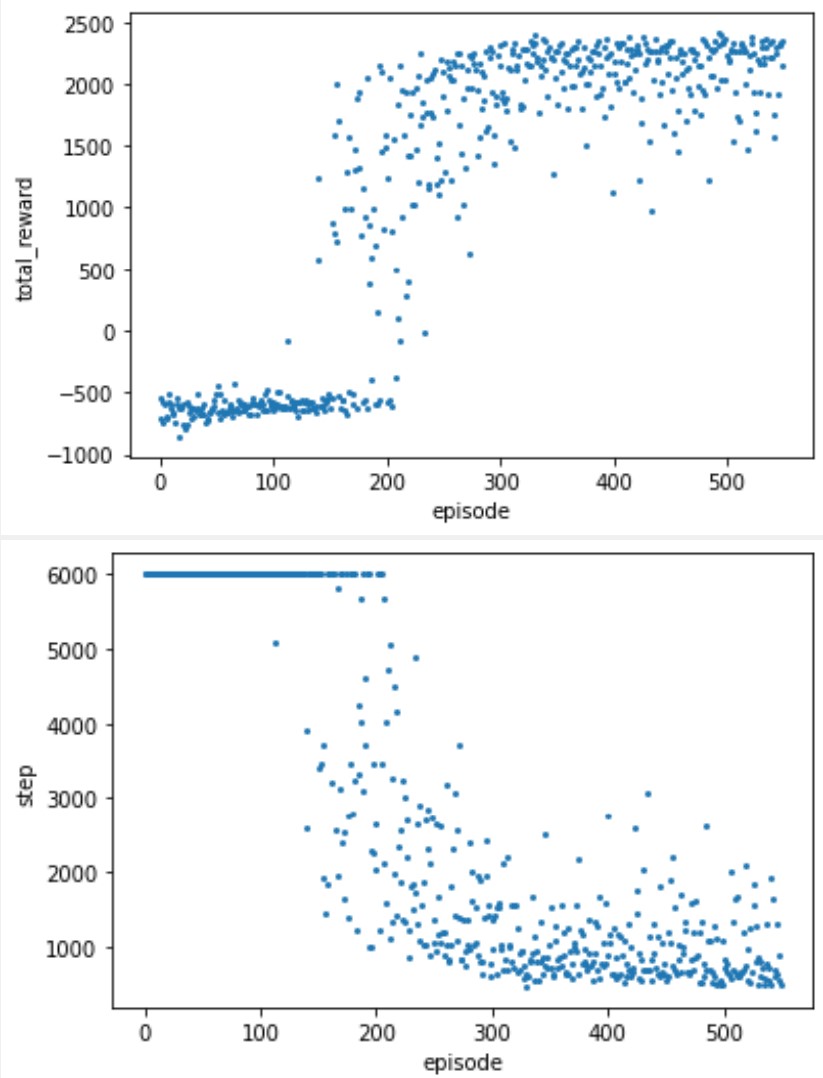}
		}	
		\caption{Overall performance in large-scale scenarios with fixed $50$ GUs, in which the maximum number of training steps per episode is set as (a) 3000 and (b) 6000.}
	\end{figure}

	\begin{figure}[htbp]
		\centering
		\subfigure[]{
			\includegraphics[width=0.4\textwidth]{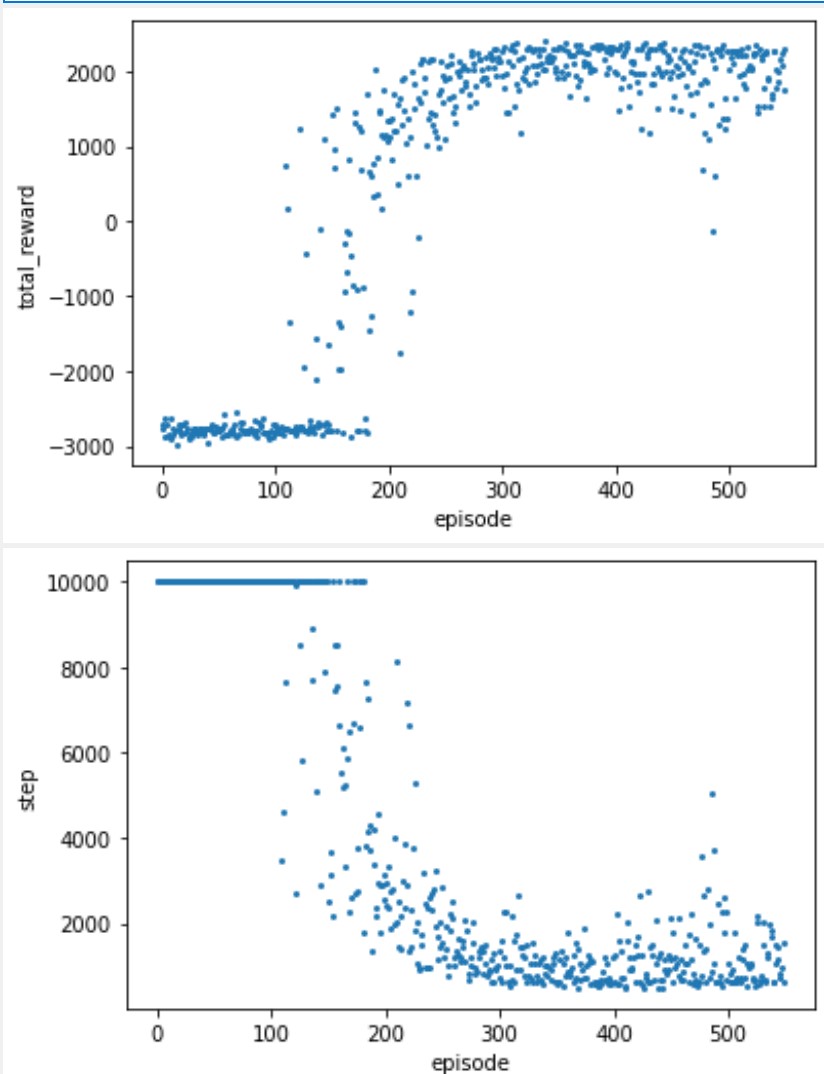}
		}	
		\subfigure[]{
			\includegraphics[width=0.4\textwidth]{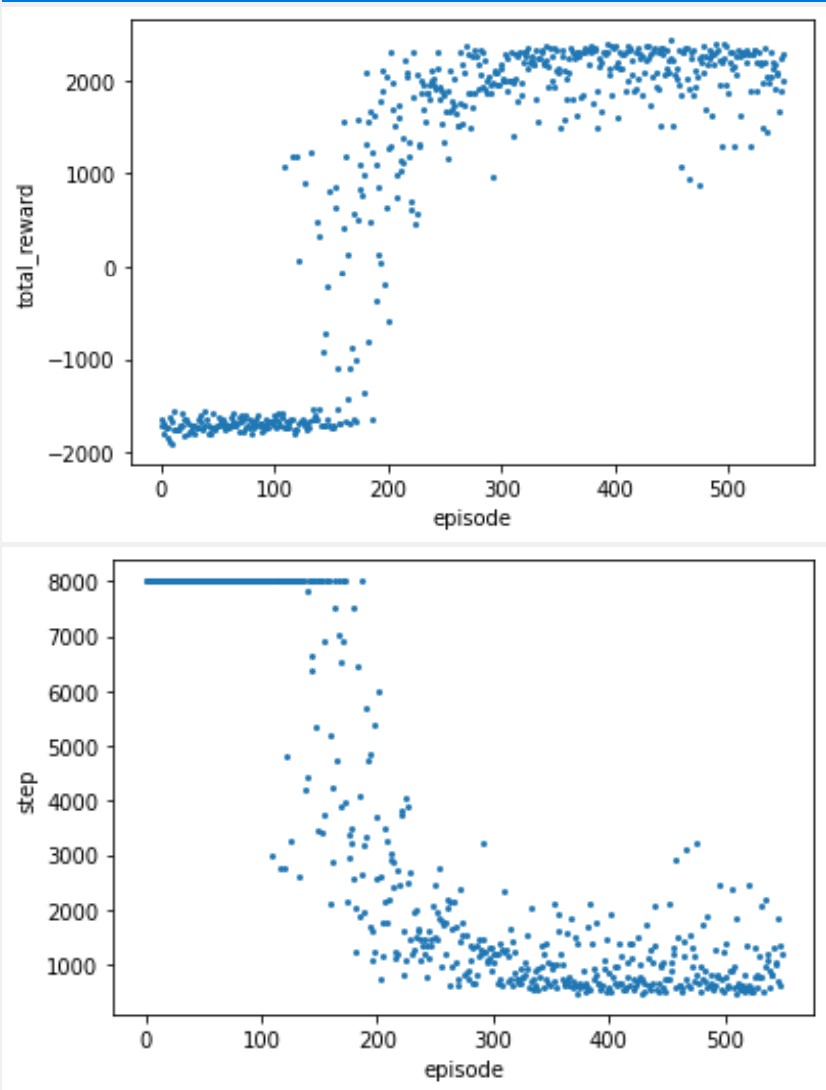}
		}	
		\caption{Overall performance in large-scale scenarios with fixed $50$ GUs, in which the maximum number of training steps per episode is set as (a) 8000 and (b) 10000.}
	\end{figure}
	
	We tested the total reward and number of steps available under multiple training parameters in a scenario with $50$ GUs, where the total reward can be seen as the overall performance that combines fairness and throughput, while the number of steps reflects the efficiency of the UAV trajectory.
	
	From the above results, even if the number of GUs is increased from $20$ to $50$, the proposed solution can still obtain good performance. We didn't compare the other two schemes because even in a $20$-GU scenario, the other two were not only unable to converge but were also significantly at a performance disadvantage.

	\begin{figure}[ht]
		\setlength{\belowcaptionskip}{3pt}
		\centering
		\includegraphics[width=0.45\textwidth]{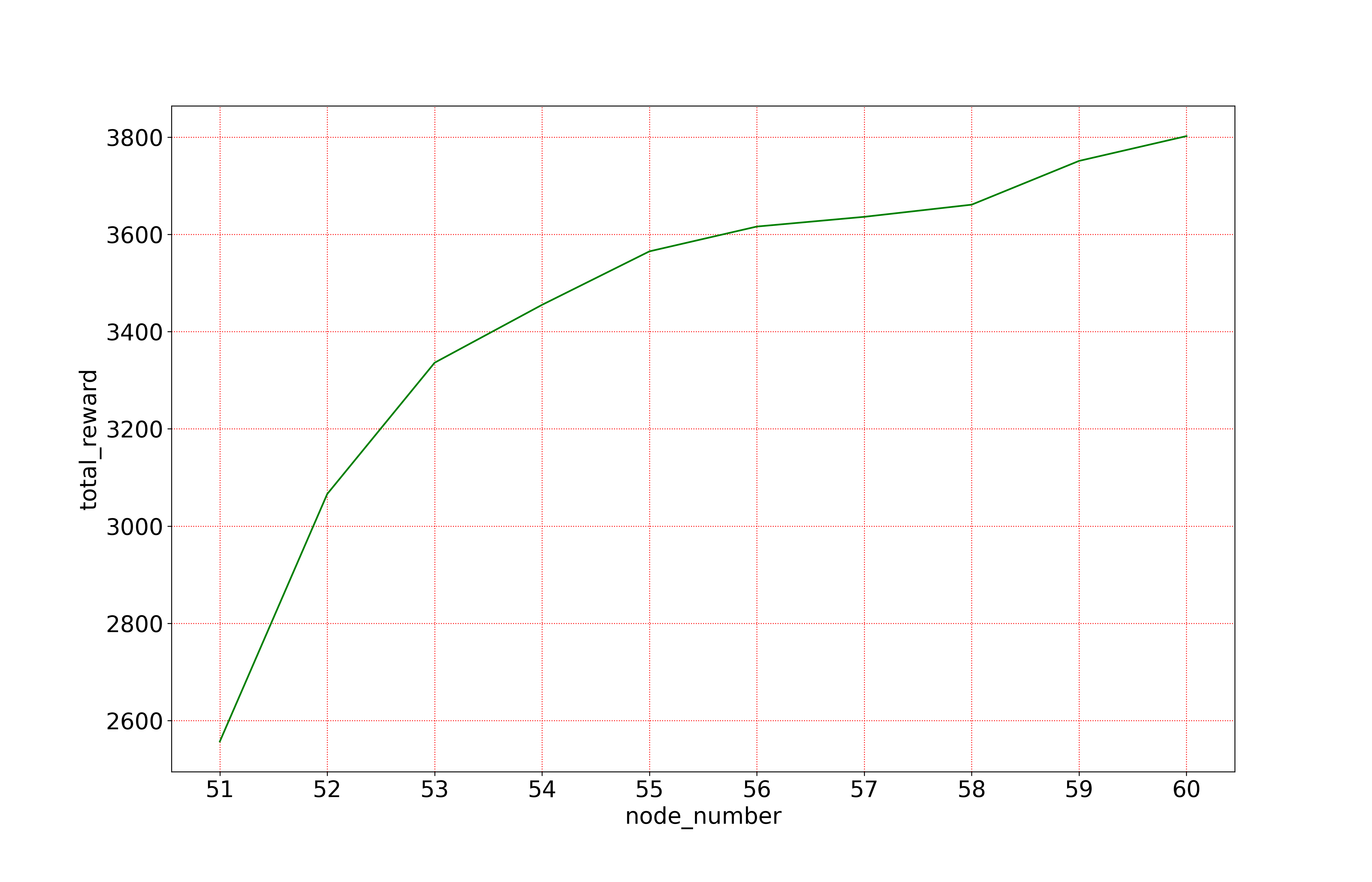}
		\caption{Overall performance in large-scale scenarios with increasing GUs}
	\end{figure}
	
	We further tested the performance of the proposed strategy when the number of GUs in the scene dynamically increased, in which only $50$ GUs are deployed in the base scene, and one new GU is deployed at the end of each training, the results are given in Fig.~3.

	\begin{figure}[ht]
		\setlength{\belowcaptionskip}{3pt}
		\centering
		\includegraphics[width=0.45\textwidth]{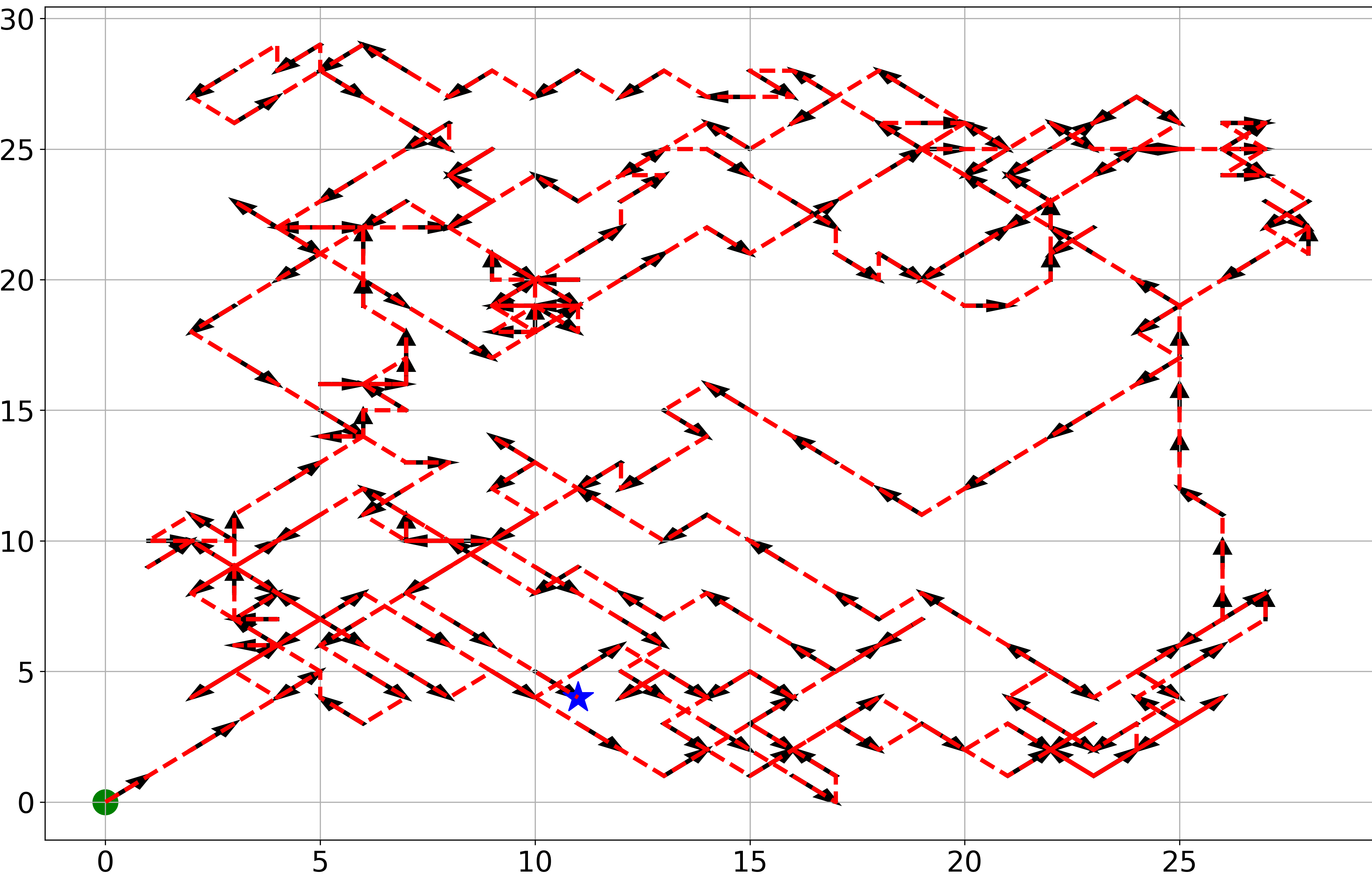}
		\caption{The trajectory of UAV in the scenario with $50$ GUs.}
	\end{figure}
	
	We further give the trajectory of the UAV in Fig.~4. The UAV starts at the green dot and ends at the blue pentagram. From the perspective of the entire trajectory, the UAV has a good effect on the coverage of the entire area, which indirectly shows that the UAV with the trend prediction function has a broader and long-term vision.
	
	\section{Notation Table}
	A table that summarizes all notations in this paper is given in Table.~1.
	\newpage
	\begin{table}
		\caption{Notations table.}
		\label{tab:1}       
		\centering 
		\begin{tabular}{lll}
			\hline\noalign{\medskip}
			Notations &\qquad\qquad\qquad\qquad Meaning  \\
			\noalign{\medskip}\hline\noalign{\medskip}
			$\Omega_{all}^t$ & The set of GUs in the scenario in time slot $t$.\\
			\noalign{\smallskip}
			$v_i^t$ & The velocity of the GU $i$ in time slot $t$.\\
			\noalign{\smallskip}
			$v_{uav}$ & The velocity of the UAV.\\
			\noalign{\smallskip}
			$\theta_i^t$ & The moving direction of GU $i$ in time slot $t$.\\
			\noalign{\smallskip}
			$h_i(t)$ & The channel coefficient between GU $i$ and UAV in time slot $t$.\\
			\noalign{\smallskip}		
			$\mathbf{L_u^t}\in\mathbb{R}^{1\times 2}$ & The location of the UAV in time slot $t$.\\
			\noalign{\smallskip}
			$\mathbf{L_i^t}\in\mathbb{R}^{1\times 2}$ & The location of GU $i$ in time slot $t$.\\		
			\noalign{\smallskip}
			$H$ & The fixed flight altitude of the UAV.\\
			\noalign{\smallskip}
			$\epsilon$ & The greedy coefficient in the selection of moving direction of GUs.\\
			\noalign{\smallskip}
			$\eta$ & The greedy coefficient in action selection of the UAV.\\
			\noalign{\smallskip}
			$p$ & The transmission power of the UAV.\\
			\noalign{\smallskip}
			$\mathbf{T_n}\in\mathbb{R}^{K\times K}$ & The $n$-th channel of the input of CNN model.\\
			\noalign{\smallskip}		
			$\tau$ & The duration of one time slot.\\
			\noalign{\smallskip}
			$\tau_c$ & The duration of the hovering time within each time slot.\\
			\noalign{\smallskip}
			$B_i^t$ & The data queue length of GU $i$ in time slot $t$.\\
			\noalign{\smallskip}
			$I_i^t$ & The newly generated data of GU $i$ in time slot $t$.\\
			\noalign{\smallskip}
			$E_f^t$ & The energy consumption of the UAV in time slot $t$.\\
			\noalign{\smallskip}
			$\underline{h}$ & The lower bound of the channel quality.\\
			\noalign{\smallskip}
			$\overline{E}$ & The upper bound of the portable energy carried by the UAV.\\
			\noalign{\smallskip}
			$f_G^t$ & The Jain's fairness index.\\
			\noalign{\smallskip}
			$\mathcal{V}_{\pi}(s_t)$ & The discounted accumulated reward from state $s_t$ to the end.\\
			\noalign{\smallskip}
			$\textbf{G}^t\in\mathbb{R}^{K\times K}$ & The matrix used to record the buffer state of GUs in the scenario.\\
			\noalign{\medskip}\hline
		\end{tabular}
	\end{table}

	\bibliographystyle{IEEEtran}

\end{document}